\documentclass{aa}

\usepackage{aabib}
\usepackage{amsmath}
\usepackage{amssymb}
\usepackage{epsfig}
\usepackage{graphicx}
\usepackage{natbib}

\begin{document} 


\abstract{We investigate the size distribution of the maximum areas and the instantaneous distribution of areas of sunspot groups using the Greenwich sunspot group record spanning the interval 1874 - 1976. Both distributions are found to be well described by log-normal functions. Using a simple model we can transform the maximum area distribution into the instantaneous area distribution if the sunspot area decay rates are also distributed log-normally. For single-valued decay rates the resulting snapshot distribution is incompatible with the observations. The current analysis therefore supports the results of \citet{Howard1992} and \citet{MartinezPillet1993}. It is not possible to distinguish between a linear and a quadratic decay law, however, with the employed data set.}

\title{On the size distribution of sunspot groups in the Greenwich sunspot record 1874-1976} 
\titlerunning{On the size distribution of sunspot groups}

\author{I. Baumann \and S.K. Solanki}
\institute{Max-Planck-Institut f\"ur Sonnensystemforschung, 37191 Katlenburg-Lindau, Germany\\
e-mail: baumann@linmpi.mpg.de}

\offprints{I. Baumann,\\ \email{baumann@linmpi.mpg.de}}

\date{\today}

\maketitle


\section{Introduction}
Sunspots appear dark on the solar surface and typically last for several days, although very large ones may live for several weeks. Sunspots are concentrations of magnetic flux, with kG magnetic field strengths. Usually, sunspots come in groups containing two sets of spots of opposite magnetic polarity.\\
The size spectrum of sunspots ranges from 3~MSH (micro solar hemispheres) for the smallest \citep{Bray1964} to more than 3\,000~MSH for very large sunspots. A quantitative study of the size distribution of sunspot umbrae has been presented by \citet{Bogdan1988}. They found a log-normal size distribution by analysing a dataset of more than 24\,000 Sunspot umbral areas determined from Mt. Wilson white-light images. Since the ratio of umbral to penumbral area depends only very slightly on the sunspot size (see the references and discussion in \citeauthor{SolankiOverview}, \citeyear{SolankiOverview}) such a distribution can be expected to be valid for sunspots as a whole.
\citet{Bogdan1988} used all sunspot observations in their sample to determine their size distribution. Since many sunspots live multiple days, the same sunspot appears multiple times in their statistics. Furthermore, in the course of its evolution, the size of a sunspot changes. Hence the method of \citet{Bogdan1988} provides the instantaneous distribution of sunspot sizes at any given time. This, however, does not in general correspond to the initial size distribution of sunspots, i.e. the distribution of freshly formed sunspots. This is expected to be very similar to the distribution of the maximum sizes of sunspots, given that sunspots grow very fast and decay slowly. For many purposes, however, the latter distribution is the more useful one. An example is when the total amount of magnetic flux appearing on the solar surface in the form of sunspots needs to be estimated (since the field strength averaged over a full sunspot is remarkably constant \citep{SolankiSchmidt1993}, the sunspot area is a good measure of the total magnetic flux).  \\
The purpose of this paper is to determine the distributions of both, the instantaneous sizes and the maximum sizes, and to compare these with each other. We determine the size distribution function of sunspot umbrae and of total sunspot areas from the digitized version of the daily sunspot observations of the Royal Greenwich Observatory (RGO).

\section{Dataset and analysis procedure}

\begin{figure}
  \resizebox{\hsize}{!}{\includegraphics{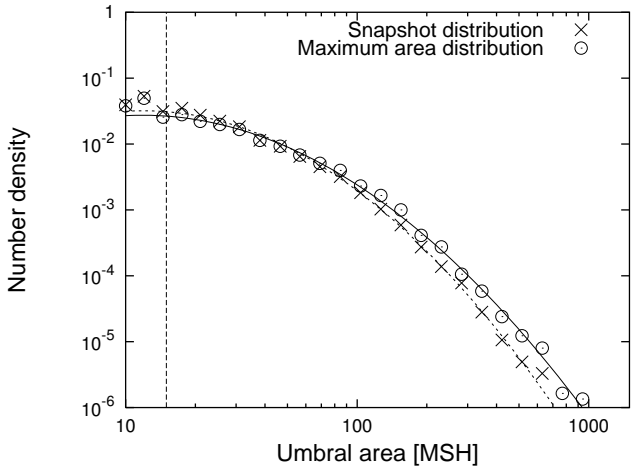}}
  \caption{Size distribution function of umbral areas obtained from the maximum development method ({\it circles}) and snapshot method ({\it crosses}). The log-normal fits are over-plotted ({\it solid line:} Fit to maximum area distribution, {\it dotted line:} Fit to snapshot distribution). The vertical line indicates the smallest umbral area considered for the fits.}
  \label{UmbralDistribution}
\end{figure}
The GPR (Greenwich Photoheliographic Results) provide the longest and most complete record of sunspot areas, spanning observations from May 1874 to the end of 1976. However, only the areas of complete sunspot groups and not of individual sunspots have been recorded. The area covered by the sunspots of a group is measured every time it is observed, i.e. every day. Besides employing these values we followed each sunspot group until it reached its maximum area. This area was stored separately. We employ in all cases true areas corrected for projection effects. \\
These stored areas can now be used to derive two different distributions of sunspot areas. If we simply form the distribution obtained from all the measured areas, we obtain the average distribution of sunspot sizes at any random instance. We call this the {\it snapshot distribution}. The snapshot distribution also underlies the study of \citet{Bogdan1988}. In general, this instantaneous size of a sunspot group will be smaller than the size of the sunspot group at its full development. In the second method, hereafter called {\it maximum development method}, the area of a sunspot group is taken at the time when the group has reached its maximum area. The maximum size is usually reached early in the development of a sunspot or sunspot group. It is followed by a steady decay \citep{McIntosh1981}.\\
The maximum group area $A_0$ determined from the Greenwich data is in general too small. Since only one observation per day is available and thus the maximum area of the spot group can be reached several hours before or after the measurement. As we consider spot groups, the different spots in the group may reach their maximum area at different times. Therefore, $A_0$ is in general somewhat smaller than the sum of the maximum areas of all the sunspots in the group. The area distribution of individual sunspots can be partly estimated by considering separately just groups of type 0, i.e. those containing just a single spot.\\
Also, visibility and projection effects lead to too small areas in the observations \citep{Kopecky1985} affecting both distributions. The RGO dataset that we use is already corrected for foreshortening. Nevertheless, in order to minimize the errors resulting from visibility corrections we use only spot groups measured within $\pm 30\,^{\circ}$ from the central meridian. When determining the maximum area of a sunspot group, we make sure that the maximum extent is reached within a longitude $\pm 30\,^{\circ}$ although the sunspot group does not necessarily have to be born within this angle.\\
We replace the continuous size distribution function $\mbox{d}N/\mbox{d}A$ by the discrete approximation $\Delta N/ \Delta A$, where $\Delta A$ is the bin width and $\Delta N$ is the raw count of the bin. Our criterion for the bin width is $20~\% $ of the geometric mean area of the bin. We include in our analysis only sunspot groups whose areas exceed a lower cut-off limit $A_{\rm min}$. For umbral areas we set the limit to $A_{\rm min}^{\rm umb} = 15$~MSH. This is similar to the cutoff of \citet{MartinezPillet1993}, which they imposed when analyzing the same data set. For total spot areas we set the cut-off limit to $A_{\rm min}^{\rm tot} = 60$~MSH. Smaller areas than $A_{\rm min}$ are not taken into account in this study, as they are falsified from enhanced intrinsic measurement errors as well as from distortions due to atmospheric seeing. \\
In order to make the size distributions for different datasets comparable, we divide $\Delta N/ \Delta A$ by the total number of spots exceeding $A_{\rm min}$. This corresponds to a normalization
\begin{figure}
  \resizebox{\hsize}{!}{\includegraphics{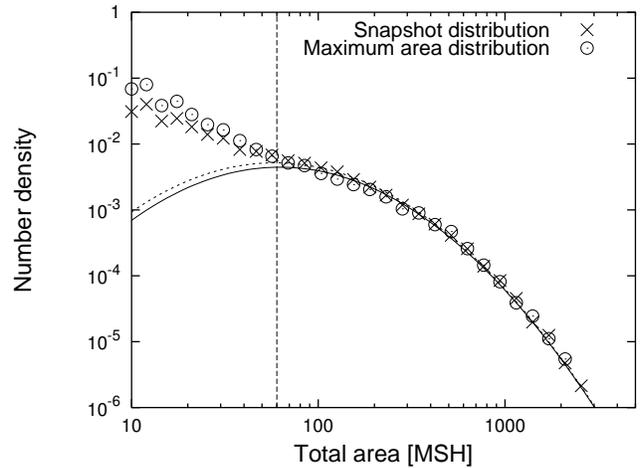}}
  \caption{Size distribution function of the total spot group areas (umbra+penumbra) obtained from the maximum development method ({\it circles}) and the snapshot method ({\it crosses}). Overplotted are the log-normal fits for $A > 60$~MSH ({\it solid line}: Maximum development method, {\it dotted line}: Snapshot method).}
  \label{TotAreaDistribution}
\end{figure}
\begin{equation}
  \int_{A_{\rm min}}^\infty \frac{\mbox{d}N}{\mbox{d}A} \mbox{d}A = 1 \, .
\label{Normalization}
\end{equation}
Finally, we fit each empirical distribution with an analytical function. In agreement with \citet{Bogdan1988} we find that a log-normal function, i.e. a continuous distribution in which the logarithm of a variable has a normal distribution, provides a good description. The general form of a log-normal distribution is
\begin{eqnarray}
  \ln \left( \frac{\mbox{d}N}{\mbox{d}A} \right)= -\frac {(\ln A - \ln \langle A \rangle)^2}{2 \ln \sigma_A} + \ln \left( \frac{\mbox{d}N}{\mbox{d}A} \right)_{\rm max} ,
\label{lognormal}
\end{eqnarray}
where $({\mbox{d}N}/{\mbox{d}A})_{\rm max}$ is the maximum value reached by the distribution, $\langle A \rangle$ is the mean area and $\sigma_A$ is a measure of the width of the log-normal distribution. Note that a log-normal function appears as a parabola in a log-log plot. \\
Log-normal distributions have been found in various fields of natural sciences. Examples are the size of silver particles in a photographic emulsion, the survival time of bacteria in disinfectants or aerosols in industrial atmospheres \citep{Crow88}, or, within solar physics, the distribution of EUV radiances in the quiet Sun \citep{Pauluhn2000}.
\begin{figure}
  \resizebox{\hsize}{!}{\includegraphics{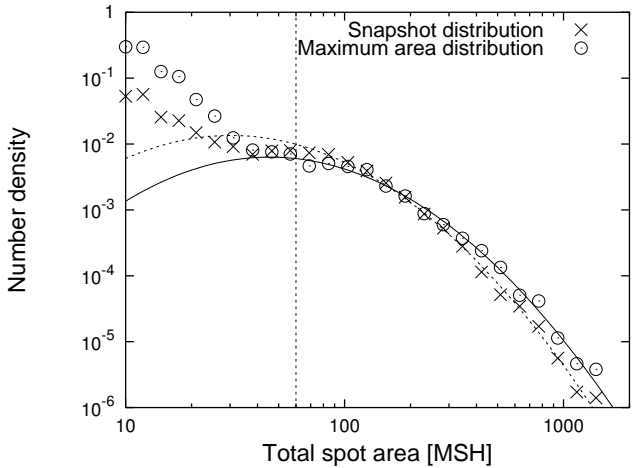}}
  \caption{Maximum area distribution ({\it circles}) and snapshot distribution ({\it crosses}) of total spot areas for single spots. Fits to the data for $A > 60$~MSH: maximum development method ({\it solid line}), snapshot method ({\it dotted line}).}
  \label{SinglesTot}
\end{figure}

\section{Results for RGO spot group areas}
\label{Comparison}

\subsection{Umbrae}
\label{Umbrae}
%
%
The size distributions of the umbral areas obtained from both, the snapshot method and the maximum development method, are shown in Fig.~\ref{UmbralDistribution}. For both methods, the resulting size distribution is well described by a log-normal function above the lower cut-off $A_{\rm min}$. As one would expect, the curve of the maximum areas lies above the snapshot curve for large sunspots. For smaller areas, the snapshot distribution is higher, resulting from the fact that the areas obtained with the snapshot method are smaller (since they include sunspots at different stages of decay), thus leading to more counts for smaller areas. The fit parameters are listed in Table~$1$. It is at first sight surprising that the size distributions obtained by both methods do not differ by a larger amount than suggested by Fig.~\ref{UmbralDistribution}. In general, the two distributions are expected to be more similar to each other if the lifetime of sunspots approaches the sampling time of the data, i.e. 1 day. For sunspots with shorter lifetimes both methods should give identical results. Therefore, the small difference between the two distributions is consistent with a relatively short average lifetime of sunspots.\\
The umbral areas for single spots from RGO are roughly a factor of $2-3$ larger than the corresponding areas from the Mt. Wilson white light plate collection. This difference probably is largely due to the fact that the RGO areas are sunspot group areas while the Mt. Wilson data analysed by \citet{Bogdan1988} give the areas of individual spots. However, since there are systematic differences also between the total areas of all the spots on a given day between data sets \citep{SolankiFligge1997,Foster2004}, other systematic differences are also likely to be present. Systematic differences lead to a shift of the RGO area distribution towards higher values of $\langle A \rangle$ and smaller values of $\sigma_A$ (Table~$1$) with respect to the Mt. Wilson dataset. The smaller value of $\sigma_A$ results from the logarithmic nature of the distribution.
%
%
\subsection{Total areas}

Fig.~\ref{TotAreaDistribution} shows the distributions for the total spot areas, i.e. the sum of umbral and penumbral area. Log-normal fits match both distributions rather well above the cut-off. However, both distributions differ even less from each other than when only the umbrae are considered (Fig.~\ref{UmbralDistribution}). Especially in the large area regime, both distributions are almost indistinguishable. Since every sunspot must have an umbra, it is not clear why the difference between the two distributions in Fig.~\ref{TotAreaDistribution} is smaller than in Fig.~\ref{UmbralDistribution}, unless it is an indication of the limits of the accuracy of the data. It may also be indicating that the decay law may be different for umbrae and sunspots as a whole.

\subsection{Total area of single spots}

\begin{figure}
  \resizebox{\hsize}{!}{\includegraphics{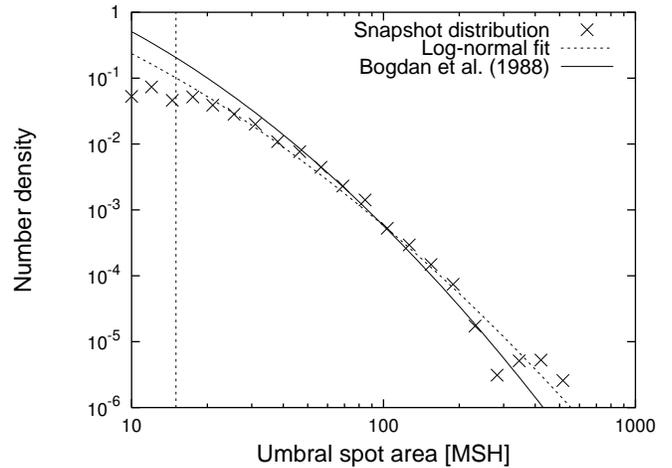}}
  \caption{Snapshot distribution of umbral spot areas for single spots ({\it crosses}), fit to the data ({\it dotted line}) and the curve from \citet{Bogdan1988} ({\it solid line}).}
  \label{SinglesUmb}
\end{figure}
In this part of the study, we extracted only Greenwich sunspot groups of type $0$, i.e. single spots (Fig.~\ref{SinglesTot}). In order to get a statistically significant dataset, we had to extend our longitudinal constraints to $\pm 60^{\circ}$ around disk center.\\ 
The difference between the snapshot and the maximum area distribution is more pronounced for total areas of single spots than for total areas of all sunspot groups. The difference in the two distributions can be explained by a similar argument as in Sect.~\ref{Umbrae}. The maximum distribution dominates for large areas, whereas the snapshot distribution shows more counts for smaller areas due to the inclusion of different decay stages of the sunspots. The similarity between Figs.~\ref{SinglesTot} and \ref{UmbralDistribution} suggests that the problem lies with Fig.~\ref{TotAreaDistribution}. It may be that when determining the total area of sunspot groups, areas of the generally short-lived pores were included in the Greenwich data set.
\subsection{Umbral areas  of single spots}
Of special interest is the snapshot distribution of umbral areas of single spots (Fig.~\ref{SinglesUmb}) because this can directly be compared to the results of \citet{Bogdan1988}. The RGO dataset displays a significantly flatter distribution than the Mt. Wilson data, i.e. the ratio of large umbrae to small umbrae is bigger for the RGO data. This systematic difference between the data sets is an indication of a systematic difference between sunspots in groups of type 0 and other spots. The parameter $\langle A \rangle$ is roughly a factor of $2$ smaller than in the corresponding Mt. Wilson data, while the width of the distribution is larger.
 \begin{table}
   \caption{Overview of the log-normal fit parameters. Due to the normalization (\ref{Normalization}) there are only two free parameters $\langle A \rangle$ and $\sigma_A$.}
   \centering
   \begin{tabular}{p{1.8cm}p{1.4cm}p{0.7cm}p{0.6cm}ll}
     \vspace{1ex}\\
     \hline
     \hline
     \vspace{0.1ex}\\
	    Data Set & Method &  $\langle A \rangle$  & $\sigma_A$ & No. of    &  Fig.\\
	             &        &                       &            & Sunspots  &      \\
	             &        &                       &            & or Groups &      \\
	      \vspace{0.1ex}\\
	      \hline
	      \vspace{0.1ex}\\
	       Mt.\;Wilson Umbrae & Bogdan et al. & 0.62 & 3.80 &  24\,615 & -\\
	      \vspace{0.1ex}\\
	       Umbrae       & Max. dev. & 11.8 & 2.55 &  3\,966 & 1\\
	       Umbrae       & Snapshot  & 12.0 & 2.24 & 31\,411 & 1\\
	      \vspace{0.1ex}\\
	       Total area   & Max. dev. & 62.2 & 2.45 &  3\,926 & 2\\
	       Total area   & Snapshot  & 58.6 & 2.49 & 34\,562 & 2\\
	      \vspace{0.1ex}\\
	       Total area & Max. dev.  & 45.5 & 2.11 &   939 & 3\\
single spots &&&\vspace{0.1cm}\\

	       Total area & Snapshot   & 30.2 & 2.14 & 15203 & 3\\
single spots &&&\\
	      \vspace{0.1ex}\\
	       Umbral area & Snapshot  &  0.27 & 6.19 &   11312 & 4\\
single spots &&&\\
	      \vspace{0.1ex}\\
	       Model        & Max. dev. & 11.8 & 2.55 &  807\,771 & 5\,a\\
	       \vspace{1mm}\\
	       Model        & Snapshot  &  &  & & \\
	               & hourly  & 7.77 & 2.80 & 21\,352\,828 & 5\,a\\
	               & daily   & 8.67 & 2.73 &  1\,092\,295 & 5\,a\\
	               & 3 days & 9.89 & 2.69 &     525\,605 & 5\,a\\
	       \vspace{0.1cm}\\
	      \hline 
   \end{tabular}
 \end{table}

\section{Modeling the snapshot distribution}

%
%
\subsection{Model description}

We have developed a simple sunspot decay model that simulates the snapshot distribution resulting from a given maximum area distribution. One aim of this modelling effort is to find out to what extend it is possible to distinguish between decay laws from the difference between the maximum area and the snapshot area distributions. Another aim is to test if, with decay laws as published in the literature, both the maximum and snapshot area distributions must have the same functional form (e.g. both be log-normally distributed).\\
We consider two kinds of maximum development distributions: a lognormal distribution (\ref{lognormal})
 and a power-law distribution of the general form
\begin{eqnarray}
 h(A)=v\cdot A^w \, .
\label{powerlaw}
\end{eqnarray}
The latter is inspired by the power-law distribution of solar bipolar magnetic regions, i.e. active and ephemeral active regions \citep{HarveyZwaan93}. \\
We assume an emergence rate of $10\,000$ spots per day. The absolute number of emerging spots does not influence the results as they are normalized (Eq.~\ref{Normalization}) and this high number is chosen in order to obtain statistically significant distributions. The constant emergence rate is a reasonable approximation of the solar case during a small period of time, i.e a few months, which is the length of time over which we let the model run. \\
 Once the spots have emerged they begin to decay immediately (the formation time of spots is short, i.e. hours \citep[e.g.][]{SolankiOverview}, and is thus neglected in the model). \\
There has been considerable debate regarding the decay law of sunspots. A number of authors have argued for a linear decay of sunspot areas with time \citep[e.g.][]{Bumba1963,MorenoInsertisVazquez1988}. Others, e.g. \cite{Petrovay1997}, found that the decay rate of a sunspot is related to its radius and thus is parabolic. The quadratic decay is also favored by models that explain the erosion of a sunspot as magnetic flux loss at the spot boundary \citep{Meyer1974}. Still others could not distinguish between a linear and a quadratic decay based on the available data \citep[e.g.][]{MartinezPillet1993}. \citet{Howard1992} and \citet{MartinezPillet1993} found that the sunspot decay rates are log-normally distributed. In view of the partly controversial situation we have computed models with all $4$ possible combinations: a) quadratic decay law with log-normally distributed decay rates, b) quadratic decay law with a single, universal decay rate, c) linear decay law with a log-normal decay rate distribution and d) linear decay law with a constant decay rate. The parabolic decay law we implement has the form
\begin{eqnarray}
  A(t)=\left(\sqrt{A_0}-\frac{D}{\sqrt{A_0}} \left(t-t_0\right)\right)^{2} \, ,
\label{DecayLaw}
\end{eqnarray}
with the added condition $A(t-t_0 > A_0/D) = 0$. The employed linear decay law has the form
\begin{eqnarray}
  A(t)=A_0-D\, (t-t_0) \, ,
\label{LinDecayLaw}
\end{eqnarray}
with $A(t-t_0 > A_0/D) = 0$. The decay rates $D$ are either given the same specified value for all sunspots in the modelled sample, or are obtained from a random number generator providing a log-normal distribution with a mean $\mu=1.75$ and a variance $\sigma^2 = 2$ following \citet{MartinezPillet1993}. \\
Combining the maximum area distribution with the decay law (Eq.~\ref{DecayLaw} or \ref{LinDecayLaw}) we can determine the resulting snapshot distribution, which can then be compared with the observed distribution. We simulate an interval of $100$~days after an initialization time of $100$ days in order to make sure that a reasonable mix of old, partly decayed spots and newly emerged spots is present. We take the fit parameters for the umbral maximum development distribution from Sect.~\ref{Comparison} as the starting distribution of our model.
\begin{figure}
 \resizebox{\hsize}{!}{\includegraphics{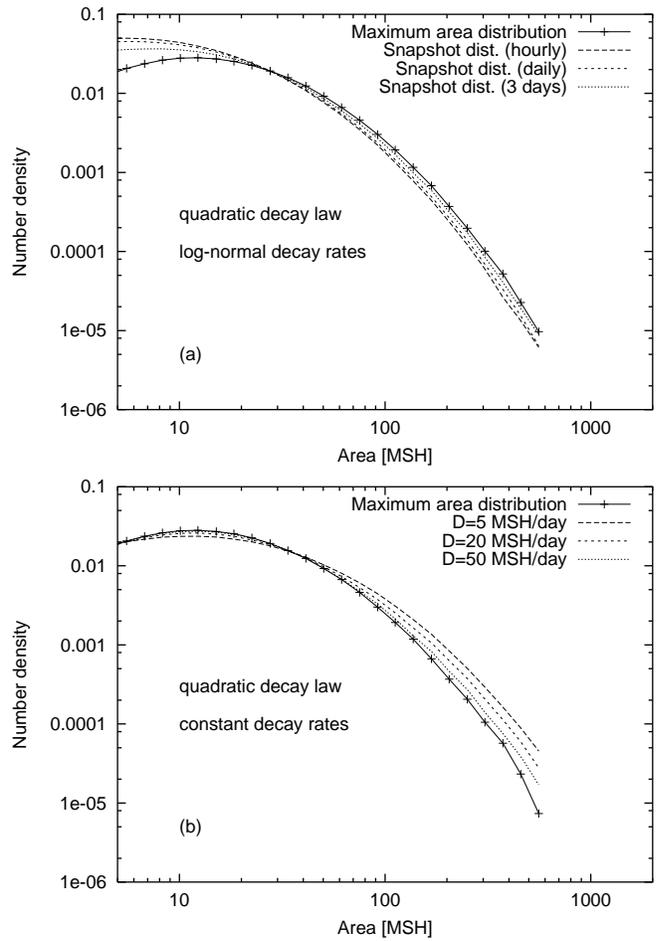}}
  \caption{Results from the model for a quadratic decay-law for (a) log-normally distributed decay rates and sampling times of $1$ hour, $1$ day and $3$ days and (b) for constant decay rates $D=5~$MSH/day, $D=20~$MSH/day and $D=50~$MSH/day and a sampling time of 1 day.}
  \label{modelQuad}
\end{figure}
\begin{figure}
 \resizebox{\hsize}{!}{\includegraphics{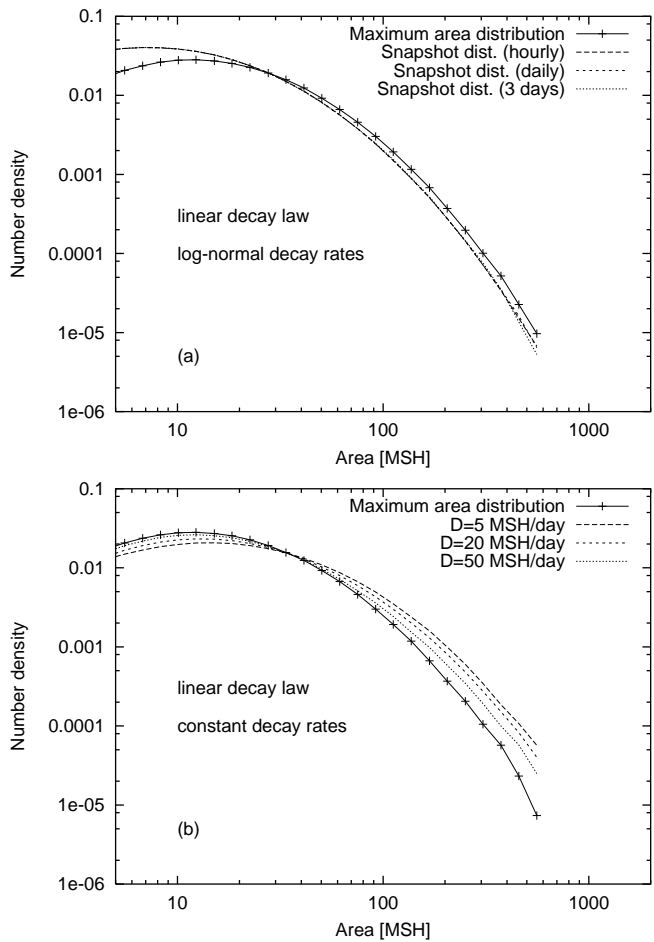}}
  \caption{Results from the model for a linear decay-law for (a) log-normally distributed decay rates and sampling times of $1$ hour, $1$ day and $3$ days and (b) for constant decay rates $D=5~$MSH/day, $D=20~$MSH/day and $D=50~$MSH/day and a sampling time of 1 day}
  \label{modelLin}
\end{figure}

\subsection{Results from the model}

%
Snapshot distributions resulting from a quadratic decay law of the form Eq.~(\ref{DecayLaw}) with log-normally distributed decay rates are plotted in Fig.~\ref{modelQuad}\,a for $3$ different sampling times. The first result is that the snapshot distributions can also be fitted well by log-normal functions. A sampling rate of $1$ day corresponds to the RGO dataset and thus can be compared with the results for umbral areas in Fig.~\ref{UmbralDistribution}. The modelled snapshot distribution matches quite well the observed snapshot distribution above the cut-off limit. At a sampling rate of $3$ days, both distributions, maximum development and snapshot, lie very close together. Such a large observing interval is comparable to the average lifetime of the spots, so that it becomes difficult to distinguish between the two distributions. For an observing frequency of 1 hour, a sampling frequency provided by the MDI dataset, the difference between the distributions is somewhat larger, as more decay stages of the spots are included in the snapshot data. When considering such a short sampling interval the formation time of the spot group becomes important and has to be taken into account, which is not included in our model.\\
%
%
In the next step, we replace the log-normally distributed decay-rates in Eq.~(\ref{DecayLaw}) by constant decay rates (Fig.~\ref{modelQuad}\,b). It is interesting that for all constant decay rates the snapshot distribution curves lie above the maximum area distribution for large sunspot areas. At first sight this appears counter-intuitive: how can the snapshot distribution show more large spots than the distribution of spot areas at maximum development? The answer lies in the normalization. For a single decay rate, small sunspots decay uniformely, so that after a given time a certain fraction has become smaller than the cut-off area and the distribution is therefore skewed towards larger spots, whose relative (but not absolute) numbers increase. The reason therefore is the normalization of the distributions. For a high decay rate (e.g. 50~MSH/day) both distribution curves lie closer together than for small decay rates (e.g. 5~MSH/day). This is understandable because a small decay rate affects more the smaller spots than the larger spots. \\
In order to see how the decay law affects the results, we repeat the above exercise, but for a linear decay law (Fig.~\ref{modelLin}). Qualitatively, a similar behaviour for both cases can be observed as in the case of a quadratic decay-law, e.g. for constant decay rates the snapshot distributions lie above the maximum area curve. When using log-normally distributed decay rates in the linear decay law (Eq.~\ref{LinDecayLaw}), the resulting snapshot curves for the three different sampling times are almost indistinguishable. We conclude from our model that it is not possible to distinguish between a linear and a quadratic decay-law by this analysis based on the Greenwich data.\\
%
%
A variability of the decay rates (log-normal distribution) thus seems necessary to yield the generally observed behaviour that the maximum area curve in general lies above the snapshot curve.\\
Finally, we check if a power-law distribution of the maximum development areas could also lead to a log-normal snapshot distribution. A power-law size distribution with an exponent $-2$ has been found by \citet{harvey93} for active regions using Kitt Peak magnetograms. Since active regions harbour sunspots, it might be worth testing if the maximum area distribution is similar to or very different from that of the host active regions. To this purpose we insert a maximum size distribution ${\mbox d}N/{\mbox d}A \sim A^{-2}$ in our model. This does not yield a log-normal snapshot distribution but rather something very close to a power-law, irrespective of the decay law. To make sure that this result is not an artefact of the special choice of the exponent of the power-law, we ran the same simulations with powers between $-1.0$ and $-3.0$. In all cases we can exclude a transformation of the power-law distribution for the maximum areas into a log-normal snapshot distribution.  \\

\section{Conclusion}

The size distribution for both, umbral and total spot area, has a pronounced, smooth log-normal shape above our lower cut-off limit. This is true for both, the instantaneous distribution of sunspot sizes (snapshot distribution) and for the distribution of sizes at the time of maximum development of each sunspot group. These two distributions are rather similar, with the snapshot distribution being slightly steeper, in general.\\
We have studied what can be learnt about sunspot decay from the comparison of these distributions, by starting from the maximum development size distribution and computing the snapshot distribution for different decay laws and parameters.\\
Both, linear and quadratic decay laws, yield qualitatively similar results, making it impossible to distinguish between them by an analysis, as carried out here. A universal decay rate of all sunspots turns out to be inconsistent with the observations, while a log-normal distribution of decay rates, as postulated by \citet{Howard1992} and \citet{MartinezPillet1993} reproduces the observations.\\
The analysis presented here can be improved with observational data that a) sample individual sunspots instead of sunspot groups, b) are observed at a higher cadence (e.g. hourly instead of daily) and c) are obtained for a homogeneous, time-independent spatial resolution. Space based imagers, such as the Michelson Doppler Imager (MDI) on the Solar and Heliospheric Observatory (SOHO) \citep{Scherrer1995} can provide such data.

\bibliographystyle{aabib} 
\bibliography{AA_2005_3415.bbl}

\begin{thebibliography}{}

\bibitem[\protect\astroncite{{Bogdan} et~al.}{1988}]{Bogdan1988}
{Bogdan} T.~J., {Gilman} P.~A., {Lerche} I., {Howard} R., 1988,
  \apj~  327, 451

\bibitem[\protect\astroncite{{Bray} \& {Loughhead}}{1964}]{Bray1964}
{Bray} R.~J., {Loughhead} R.~E., 1964,
\newblock {Sunspots},
\newblock Chapman \& Hall, London

\bibitem[\protect\astroncite{{Bumba}}{1963}]{Bumba1963}
{Bumba} V., 1963,
  Bull. Astron. Inst. Czechoslovakia~  14, 91

\bibitem[\protect\astroncite{{Crow} \& {Shimizu}}{1988}]{Crow88}
{Crow} E.~L., {Shimizu} K.~E., 1988,
\newblock Lognormal Distributions: Theory and Applications,
\newblock Dekker,
  New York

\bibitem[\protect\astroncite{{Fligge} \& {Solanki}}{1997}]{SolankiFligge1997}
{Fligge} M., {Solanki} S.~K., 1997,
  \solphys~  173, 427

\bibitem[\protect\astroncite{Foster}{2004}]{Foster2004}
Foster S., 2004,
\newblock Ph.D. thesis,
  {University of Southhampton, Faculty of Science, School of Physics and
  Astronomy}

\bibitem[\protect\astroncite{{Harvey}}{1993}]{harvey93}
{Harvey} K., 1993,
\newblock Ph.D. thesis,
  Astron. Inst. Utrecht Univ.

\bibitem[\protect\astroncite{{Harvey} \& {Zwaan}}{1993}]{HarveyZwaan93}
{Harvey} K.~L., {Zwaan} C., 1993,
  \solphys~  148, 85

\bibitem[\protect\astroncite{{Howard}}{1992}]{Howard1992}
{Howard} R.~F., 1992,
  \solphys~  137, 51

\bibitem[\protect\astroncite{{Kopeck\'y} et~al.}{1985}]{Kopecky1985}
{Kopeck\'y} M., {Kuklin} G.~V., {Starkova} I.~P., 1985,
  Bull. Astron. Inst. Czechoslovakia~  36, 189

\bibitem[\protect\astroncite{{Mart\'inez Pillet}
  et~al.}{1993}]{MartinezPillet1993}
{Mart\'inez Pillet} V., {Moreno-Insertis} F., {V\'azquez} M., 1993,
  \aap~  274, 521

\bibitem[\protect\astroncite{{McIntosh}}{1981}]{McIntosh1981}
{McIntosh} P.~S., 1981,
\newblock in The Physics of Sunspots, L.~Cram, J.H.~Thomas (Eds.),
  NSO, Sunspot, N.M.,
  p.~7

\bibitem[\protect\astroncite{{Meyer} et~al.}{1974}]{Meyer1974}
{Meyer} F., {Schmidt} H.~U., {Wilson} P.~R., {Weiss} N.~O., 1974,
  \mnras~  169, 35

\bibitem[\protect\astroncite{{Moreno-Insertis} \&
  {V\'{a}zquez}}{1988}]{MorenoInsertisVazquez1988}
{Moreno-Insertis} F., {V\'{a}zquez} M., 1988,
  \aap~  205, 289

\bibitem[\protect\astroncite{{Pauluhn} et~al.}{2000}]{Pauluhn2000}
{Pauluhn} A., {Solanki} S.~K., {R{\" u}edi} I., {Landi} E., {Sch{\" u}hle} U.,
  2000,
  \aap~  362, 737

\bibitem[\protect\astroncite{{Petrovay} \& {van
  Driel-Gesztelyi}}{1997}]{Petrovay1997}
{Petrovay} K., {van Driel-Gesztelyi} L., 1997,
  \solphys~  176, 249

\bibitem[\protect\astroncite{{Scherrer} et~al.}{1995}]{Scherrer1995}
{Scherrer} P.~H., {Bogart} R.~S., {Bush} R.~I., {Hoeksema} J.~T., {Kosovichev}
  A.~G., {Schou} J., {Rosenberg} W., {Springer} L., {Tarbell} T.~D., {Title}
  A., {Wolfson} C.~J., {Zayer} I., {MDI Engineering Team}, 1995,
  \solphys~  162, 129

\bibitem[\protect\astroncite{{Solanki}}{2003}]{SolankiOverview}
{Solanki} S.~K., 2003,
  \aapr~  11, 153

\bibitem[\protect\astroncite{{Solanki} \& {Schmidt}}{1993}]{SolankiSchmidt1993}
{Solanki} S.~K., {Schmidt} H.~U., 1993,
  \aap~  267, 287

\end{thebibliography}

\end{document}